# Elemental Analysis of Glass and Bakelite Electrodes Using PIXE Facility

Manisha[1]*, V. Bhatnagar[1], J. S. Shahi[1], S. Verma[1],
B. P. Mohanty[2], A. Kumar[1]

[1]*Department of Physics, Panjab University, Chandigarh-160014, India.*
[2]*Department of Biophysics, Panjab University, Chandigarh-160014, India.*

*Corresponding author : manisha.lohan@cern.ch

**Abstract :** The evolution of particle detectors dates back to the discovery of X-rays and radioactivity in 1890s. In detector history, the Resistive Plate Chambers (RPCs) are introduced in early 1980s. An RPC is a gaseous detector made up of two parallel electrodes having high resistivity like that of glass and bakelite. Currently several high energy physics experiments are using RPC-based detector system due to robustness and simplicity of construction. In each and every experiment, RPCs have to run continuously for several years. So, it demands an indepth characterization of the electrode materials. In the present study, an elemental analysis of locally available glass and bakelite samples is done using PIXE facility available at Panjab University Cyclotron, Chandigarh. PIXE measurements are done using 2.7 MeV proton beam incident on the electrode sample target. The constituent elements present in these electrode samples are reported.

Keywords: Proton Induced X-ray Emission (PIXE), Glass, Bakelite.

# 1 . Introduction

An RPC is a gaseous detector utilising a constant and unifrom elecrtic field produced by two highly resistive electrode plates like glass and bakelite [1]. These characteristics make both, glass and bakelite RPCs active candidates in various running as well as future HEP experiments for different applications [2]. In STAR experiment at RHIC [3]; ALICE, ATLAS and CMS experiments at LHC [4–6]; Belle experiment at KEK are using glass and bakelite RPC based detector systems [7]. The proposed Iron CALorimeter (ICAL) detector at the underground India-based Neutrino Observatory (INO) facility is also planning to use glass RPCs as the active detector elements [8]. For long life time of experiment, it is necessary to do elaborate study to characterize electrode materials. In reported study, elemental analysis of glass and bakelite samples is done using PIXE facility available at Cyclotron laboratory [9], Chandigarh. Elemental analysis gives an insight in to the elemental composition of glass and bakelite electrodes; glass and bakelite electrode samples procured from local and international market are found to be having similar elemental composition. Therefore, locally available electrode samples could be used.

# 2. PIXE

PIXE is a well established analytical technique of X-ray spectroscopy, which is used for rapid and simultaneous multielement analysis. To perform PIXE of a sample, it is bombarded with high energy charged particles, which causes ionization (due to coulomb interaction) of inner shell electrons, electrons from higher shell fill that vacancy in inner shell and difference of binding energies of two shells is emitted in the form of X-rays. Identification of constituent elements of analyzed sample is done on the basis of wavelength of emitted characteristic X-rays and concentration is predicted from intensities of characteristic X-rays. These characteristic X-rays are detected by high purity germanium (HPGe) X- ray detector-GUL0035 [9].

## 2.1 Experimental setup and data analysis

PIXE measurements are done using 2.7 MeV proton beam, incident on target (sample to be analyzed) using a graphite collimator having 1 mm diameter. Al absorbers are used to improve count rate of high Z elements (Z>34). Emitted X-rays are detected by HPGe X-ray detector, positioned at an angle of 45° to the incident proton beam axis. PIXE spectrum is obtained with the help of MAESTRO (for windows OS) program and analyzed with GUPIX software [10].

# 3 . Results

PIXE spectrum of glass and bakelite samples are shown in Figure 1. Ninteen elements Si, P, S, K, Ca, Sc, Ti, Mn, Fe, Co, Ni, Cu, Zn, Sn, Sr, Cl, Cr, Nb and Sb are predicted from PIXE measurements of glass and bakelite. Concentration of predicted elements is shown in Table 1.

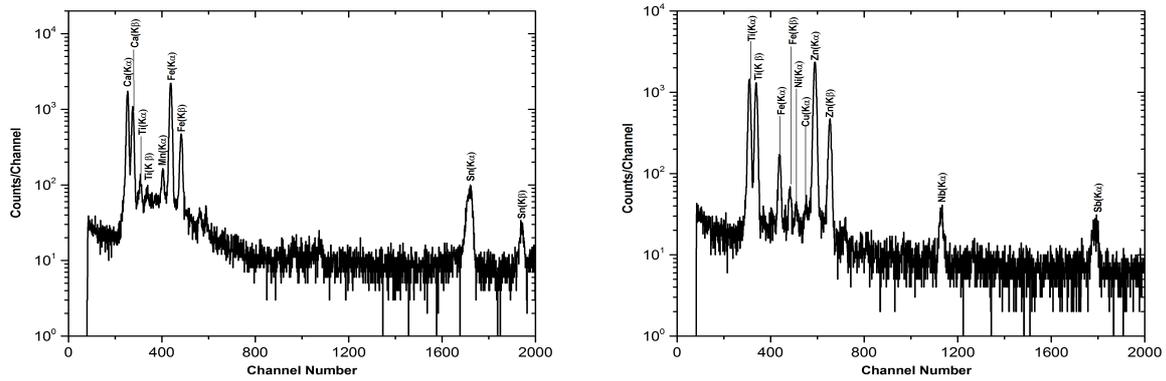

**Figure 1. PIXE spectrum of glass (left) and bakelite (right).**

Sample codes: A: Asahi; B: Modi; C: Saint Gobain; D: IEL; E: Hylam; F: Italian; G: FormicaW; H: FormicaB.

**Table 1. Elemental concentrations of glass and bakelite samples predicted from PIXE measurements.**

| Element | A(PPM) | B(PPM) | C(PPM) | D(PPM) | E(PPM) | F(PPM) | G(PPM) | H(PPM) |
|---|---|---|---|---|---|---|---|---|
| Si | 449221 | 465556 | 743768 | 914.3 | 2635 | 344 | - | 14883 |
| P | 698 | - | 940.6 | 195.6 | 111.7 | 448 | 1402 | - |
| Ca | 51499 | 52938.7 | 83120 | 10587 | 303 | 771 | 663.6 | 3966 |
| Sc | 228.2 | 648.9 | 1056.4 | 3312 | 29.5 | 8.6 | 1037.1 | 99.2 |
| K | 1292.7 | 1444.2 | 6116.9 | 248.3 | 1120 | 129 | 87.1 | 190 |
| S | 985 | 875.6 | 1542.3 | 735.2 | 180 | 2051 | 426.5 | 53051 |
| Cl | - | - | - | 2137 | 3163 | 299 | 285.4 | 2995 |
| Fe | 976.3 | 610.9 | 1751.1 | 321.5 | 282 | 84.1 | 268.4 | 283.4 |
| Ti | 121.4 | 150.5 | 193 | 258 | 68432.5 | 1.1 | 121531 | 1897.2 |
| Mn | 37.7 | 63.1 | 60.5 | 26.7 | - | 30.3 | - | 841.3 |
| Cr | - | - | - | 43.5 | 131.2 | 21.8 | 593.3 | - |
| Zn | - | 7.2 | 6.7 | 1968.7 | 1119 | 5.7 | 34.6 | 11.4 |
| Sn | - | 3413.3 | 2848.9 | - | - | - | - | - |
| Co | 13.2 | 10.8 | 42.7 | 109.6 | 15.4 | 1.6 | 4.6 | - |
| Ni | 3.8 | - | 6.3 | 997.4 | 11 | 3.9 | 10.2 | 6.3 |
| Cu | - | - | - | 65.6 | 91 | 2.3 | 66.3 | 46.1 |
| Sr | 25.3 | 13.3 | 95.6 | - | - | - | - | - |
| Nb | - | - | - | 1.3 | 51.5 | 2.2 | 1.6 | 1.1 |
| Sb | - | - | - | - | 310 | - | - | - |

# 4